\documentclass[letterpaper]{article}

\usepackage[T1]{fontenc}

\usepackage{geometry}
\usepackage{setspace}

\usepackage{achemso}

\usepackage{graphicx}
\usepackage{float}
\usepackage{amssymb}
\usepackage{amsmath}
\usepackage{textcomp}
\usepackage{epstopdf}
\usepackage[dvipsnames]{xcolor}
\usepackage{units}
\usepackage[english]{babel}
\newcommand*{\eg}{e.\,g.,}
\newcommand*{\ie}{i.\,e.,}

\newcommand*{\etal}{\textit{et~al.}}
\usepackage{blindtext}
\usepackage{blindtext}
\usepackage{ulem}
\newfloat{scheme}{htbp}{los}
\floatname{scheme}{Scheme}
\floatname{chart}{Chart}
\newfloat{graph}{htbp}{loh}
\usepackage{soul}

\usepackage{chemformula} 
\usepackage[version = 4]{mhchem} 

\usepackage{authblk}
\author[1,2,3,*]{H.~Hübschmann}
\author[1,2,3]{G.~Berth}
\author[4]{M.~Groll}
\author[5]{K.~Burgholzer}
\author[5]{A.~Bonanni}
\author[1,2,3]{K.~D.~Jöns}
\author[1,3]{U.~Gerstmann}
\author[1,3]{W.~G.~Schmidt}
\author[1,2,3,$\dagger$]{A.~Bocchini}

\affil[1]{Department of Physics, Paderborn University, 33095 Paderborn, Germany}
\affil[2]{Institute for Photonic Quantum Systems (PhoQS), Paderborn University, 33095 Paderborn, Germany}
\affil[3]{Center for Optoelectronics and Photonics Paderborn (CeOPP), Paderborn University, 33095 Paderborn, Germany}
\affil[4]{Chair of Materials Science (LWK), Paderborn University, 33095 Paderborn, Germany}
\affil[5]{Institute of Semiconductor and Solid State Physics, Johannes Kepler University Linz, 4040 Linz, Austria}

\title{Surface-exciton enhanced SHG response in few-layer 2H-TMDC}
\date{*Email: henry.huebschmann@upb.de\\
      $\dagger$Email: adriana.bocchini@upb.de}

\usepackage{todonotes}
\usepackage{wrapfig}
\usepackage{graphicx}

\usepackage[draft,todonotes={textsize=tiny}]{changes} 
\definechangesauthor[name=Adriana, color=TealBlue]{AB}

\begin{document}

\maketitle

\begin{abstract}
\begin{minipage}[l]{0.38\textwidth}
We explore the nonlinear optical properties of few-layer MoS$_2$ by means of polarization- and laser-power-dependent measurements as well as \textit{ab initio} techniques.
While for even-layer samples a weak second-harmonic (SH) signal can be attributed to the presence of surface defects or interface effects, our measurements resolve a layer-number dependent signal for odd-layer samples.
For the excitation energy of $780$\,nm, we find that the SH intensity decreases steadily with the layer number.
Our simulations demonstrate that
\end{minipage}%
\vspace*{0.7mm}
\begin{minipage}[l]{0.62\textwidth}
\includegraphics[width=3.25in, height=1.75in]{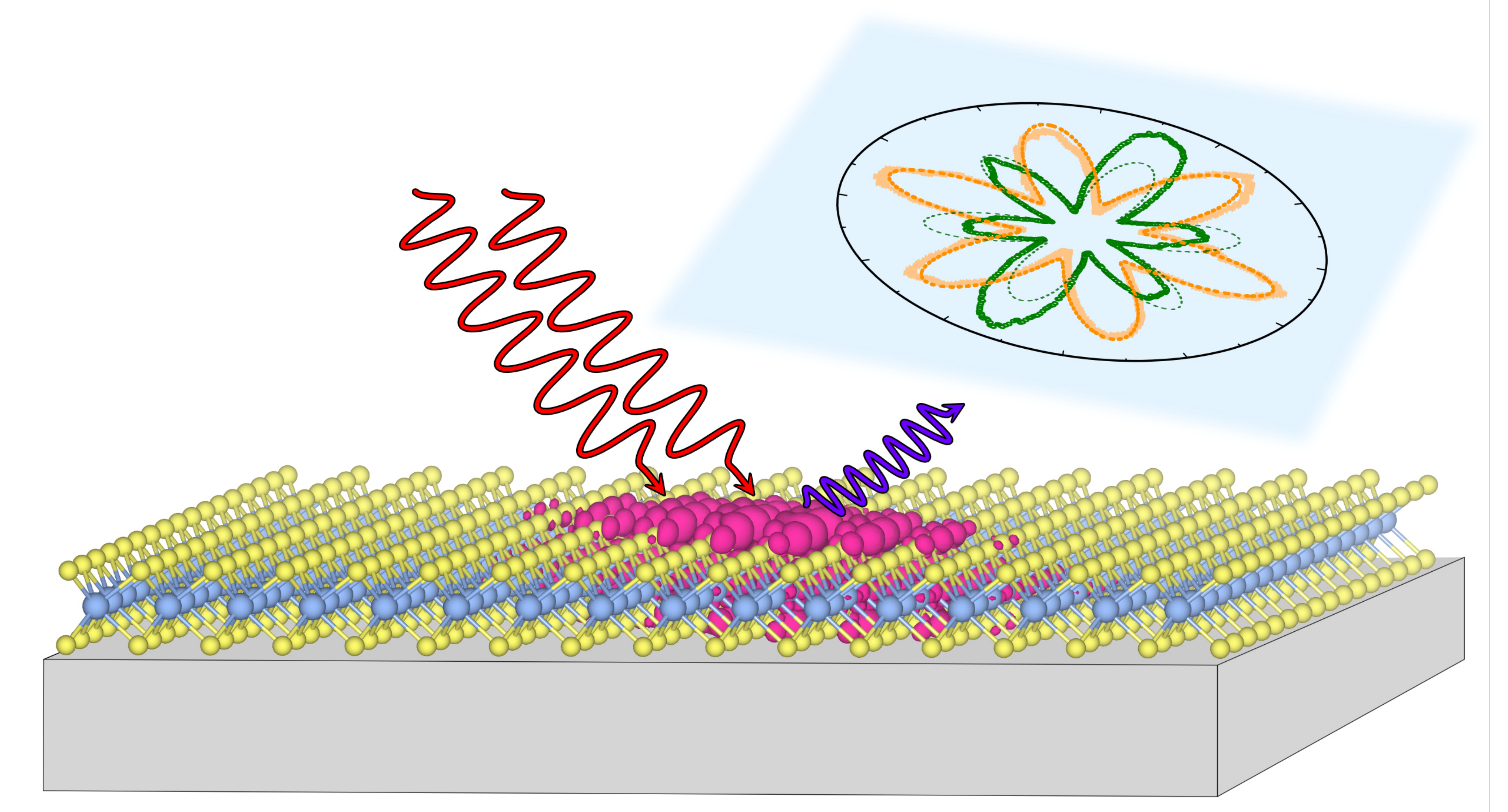} 
\end{minipage}%

\begin{minipage}[l]{0.88\textwidth}
  this effect cannot be purely attributed to modifications of the band structure, but requires the inclusion of excitonic effects and can be explained by the increasing delocalization of excitons with increasing sample thickness.
\end{minipage}%
\end{abstract}

\section*{Keywords}
MoS$_2$, Second harmonic generation, SHG, Excitonic effects, 2H-TMDC

\section*{List of Abbreviations}
\begin{tabular}{l l}
   $\bullet$ Atomic force microscopy & AFM \\
   $\bullet$ Continuous wave laser & CW laser \\
   $\bullet$ High-angle annular dark-field & HAADF \\
   $\bullet$ Molybdenum & Mo \\
   $\bullet$ Molybdenum disulfide &  MoS$_2$ \\
   $\bullet$ Monolayer & ML \\
   $\bullet$ Scanning transmission electron microscopy & STEM \\
   $\bullet$ Second-harmonic & SH \\
   $\bullet$ Silicon & Si \\
   $\bullet$ Silicon oxide & SiO$_2$ \\
   $\bullet$ Transition metal dichalcogenides & TMDCs \\

\end{tabular}

\section{Introduction}

Atomically thin-layered transition metal dichalcogenides (TMDCs), \eg\ molybdenum disulfide (MoS$_2$), have been studied extensively over the last two decades \cite{pakdel2012, liu2011, mahatha2012, kim2012, ci2010, Song2012}. 
They show interesting properties like strong light-matter interaction, tunable electronic structures as well as high integrability into photonic and/or electronic devices \cite{Li2015, Cheng2014, Akinwande2014, Lee2014, Kumar2025, Tan2020, Coskun2025, Abajo2025, Liu2024}. 
Here, major attention is given to the investigation of excitonic effects on the optical properties of thin TMDCs \cite{Qiu2013, Ramasubramaniam2012, Cheiwchanchamnangij2012, Berkelbach2013, Komsa2012, He2014, Latini2015, Hueser2013, Berghaeuser2014, Fugallo2021, Yao2026}.
In addition, the observation of single photon emission in such materials opened the utilization in quantum technology by, e.g., strain engineering \cite{Shiue2017, Guo2020, Esposito2025, Gu2025, Thayil2025, Gupta2023, Turunen2022}. 
Another promising research direction focuses on nonlinear optics \cite{Autere2018}. 
The extraordinary nonlinear response of these systems, in fact, is not only interesting for applications, but also for fundamental studies \cite{Li2013, Soh2018, malard2013, Shree2021, Gruening2014, kumar2013, Hung2023}:
The general crystal symmetry depends on the exact layer number as well as the stacking sequence \cite{Zhao2016}, whereby the so-called 2H stacking preserves the bulk inversion symmetry and therefore only allows for second harmonic generation for odd layer numbers \cite{Li2013}.
Consequently, for a knowledge-based  implementation of such structures into devices, the fundamental behavior needs to be understood correctly. 

The wavelength-dependent nonlinear response of 2D-MoS$_2$ has already been investigated, and it is found that absorption effects as well as excitonic features modify the strength of the nonlinear coefficient $\chi^{(2)}$ greatly \cite{Shree2021, Trolle2015, trolle2014, Seyler2015, Wang2015, Wang2015a, Glazov2017, Skachkov2026, Ruan2024}.
Here, we go beyond these previous studies and explore the combined influence of film thickness and excitonic effects on the nonlinear optical response of few-layer MoS$_2$ systems by combining experiment with \textit{ab initio} theory.

\section{Methodology}
\subsection{Experimental}

Exfoliated few-layer MoS$_2$ of up to 9 monolayers (ML) in thickness on SiO$_2$/Si substrate are investigated.
Since a precise knowledge of the exact layer number in the samples is crucial for the findings in this study, we perform a series of measurements to pre-characterize the structures:
For this, we utilize a high-resolution Keyence VHX-7000 optical microscope to determine the sample thickness via optical contrast calibration. Additionally, optical control enables the inspection and locating of layered structures for further analysis.
In addition, a Raman analysis is performed, which is commonly applied to determine the layer number in few-layer configurations.
Here, a $532$\,nm CW laser at incident power lower $1$\,mW is used.

Furthermore, the 2H-stacking of the samples is confirmed by means of scanning transmission electron microscopy (STEM) using a probe-side C$_{\mathrm{S}}$-corrected microscope (JEOL JEMARM 200F) at an acceleration voltage of 80\,keV and a semi-convergence angle of 30\,mrad. For the investigation of the atomic structure, a high-angle annular dark-field (HAADF) detector which covers a collection angle interval of 53--191\,mrad is employed. The MoS$_2$ flakes were transferred to silicon nitride window grids (Pelco\textsuperscript{\textregistered} from Plano GmbH), covered with a 200\,nm thick holes SiN membrane by a PDMS-based transfer process\cite{Castellanos2014, Schranghamer2021}, enabling the high-resolution STEM investigations of free-standing MoS$_2$ mono- and bilayers.

Experimental power-controlled, polarization-dependent second-harmonic-generation (SHG) measurements are conducted in back-scattering direction.
During all SH measurements, sample positioning with respect to the laser focus is realized by a 3D-positioning stage (PI Nanocube).
Thereby, we use a pulsed Toptica (FFSmart 780) fiber laser centered at 780\,nm with a repetition rate of 80\,MHz and a pulse width of 100\,fs for excitation.
The light polarization is determined via a monotorized half-wave plate, while a dichroic beam splitter and color filters are involved for spectral filtering to ensure that all detected photons originate from SH processes in the material itself.
Upon polarization analysis and spectral filtering, an infinite-corrected objective (numerical aperture of 0.95 and magnification of 100$\times$) is applied to both focus the excitation radiation on the MoS$_2$ few-layer samples and to collect the SH radiation.
Finally, the generated SH light is detected using an avalanche photo diode.

\subsection{Theory\label{sec:methods_dft}}

Ground-state geometries and electronic structures of few-layer MoS$_2$ are calculated within density functional theory (DFT) as implemented in the open-source package \textsc{Quantum ESPRESSO} \cite{Giannozzi2009, Giannozzi2017}.
To model thin-film MoS$_2$, $10$\,\AA\ of vacuum are added on top of up to five monolayers of MoS$_2$ along the $[001]$ crystal direction following the 2H stacking. 
During geometry relaxation, the lattice constants $a=b$ are kept fixed at their experimental value \cite{fan2014}.
The atomic positions, on the other hand, are optimized until fluctuations in total energy and forces are below $10^{-4}$\,Ry and $10^{-8}$\,Ry/Bohr, respectively.
Here, an energy cutoff of 90\,Ry is used for the plane-wave expansion, and the surface Brillouin zone is sampled using a shifted, $12 \times 12$ \textbf{\textit{k}}-point mesh.

Van-der-Waals (vdw) effects are taken into account via the semi-empirical Grimme-D3 scheme \cite{grimmed3}.
The effects of spin-orbit coupling (SOC) onto the second-order tensor $\chi^2_{yyy}$ have been shown to yield a splitting of the first peak, \ie\ the $\nicefrac{A}{2}$ and $\nicefrac{B}{2}$ exciton peaks caused by direct transitions at the $K$ point \cite{trolle2014,jiang2022}.
The SOC-related splitting of the main exciton peak $\nicefrac{C}{2}$, on the other hand, which is caused by transitions between the $\Gamma$ and the $K$ point, is far less pronounced \cite{trolle2014, jiang2022}. 
Therefore and to reduce the computational costs, especially in combination with higher levels of theory and larger system sizes, the SOC effects are neglected in this study.

The electron-ion interactions are included via norm-conserving pseudopotentials, and for the electron exchange and correlation (XC) effects, the generalized gradient approximation (GGA), through the Perdew-Burke-Ernzerhof functional (PBE) \cite{pbe}, is applied.
Taking advantage of error cancellation, this functional, in fact, allows for a realistic description of the optical band-gap width of bulk MoS$_2$, \ie\ $E_{\mathrm{K}}^{\mathrm{PBE}} = 1.63$\,eV vs.~$E_{\mathrm{K},v1} = 1.88$\,eV and $E_{\mathrm{K},v2} = 2.06$\,eV \cite{beal1979}.

For the nonlinear optical properties, we use the \textsc{Yambo} code \cite{yambo2009, yambo2019}, allowing for their computation by means of the dynamical Berry-phase approach \cite{lumen}.
Here, we find that a cutoff energy of $150$\,Ry as well as a $21 \times 21$ $\Gamma$-centered \textbf{\textit{k}}-point mesh yield satisfactorily converged spectra, while keeping the computational cost affordable.
Within the independent particle approximation (IPA), SH spectra are obtained using 13 bands (\ie\ seven filled and five empty bands) per ML in the system.
In addition, quasiparticle as well as excitonic effects are taken into account via the \textit{G$_0$W$_0$} approximation and following the time-dependent screened Hartree-Fock approach \cite{lumen,lumenMoS2} (TD-SHF), respectively.
Here, we calculate the Coulomb screening involving 40 bands per ML, while the SH spectra are obtained using six valence and four conduction bands per ML.

\section{Results and discussion\label{sec:results}}

\begin{figure}[!ttt]
\centering
\includegraphics[width=0.42\textwidth]{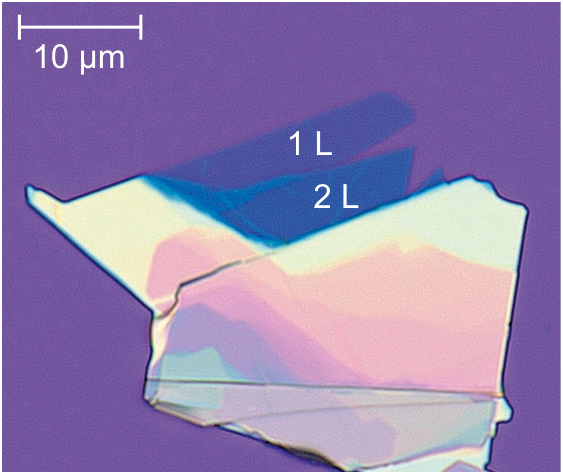}
    \vspace*{-1mm}
    \caption{Micrograph of a MoS$_2$ flake on SiO$_2$/Si-substrate showing a mono- and a bilayer in thickness. 
    Note that darker regions correspond to thinner areas.
    }
    \label{fig:SI-micrograph}
\end{figure}

\begin{figure}[!bbb]
\centering
\includegraphics[width=0.48\textwidth]{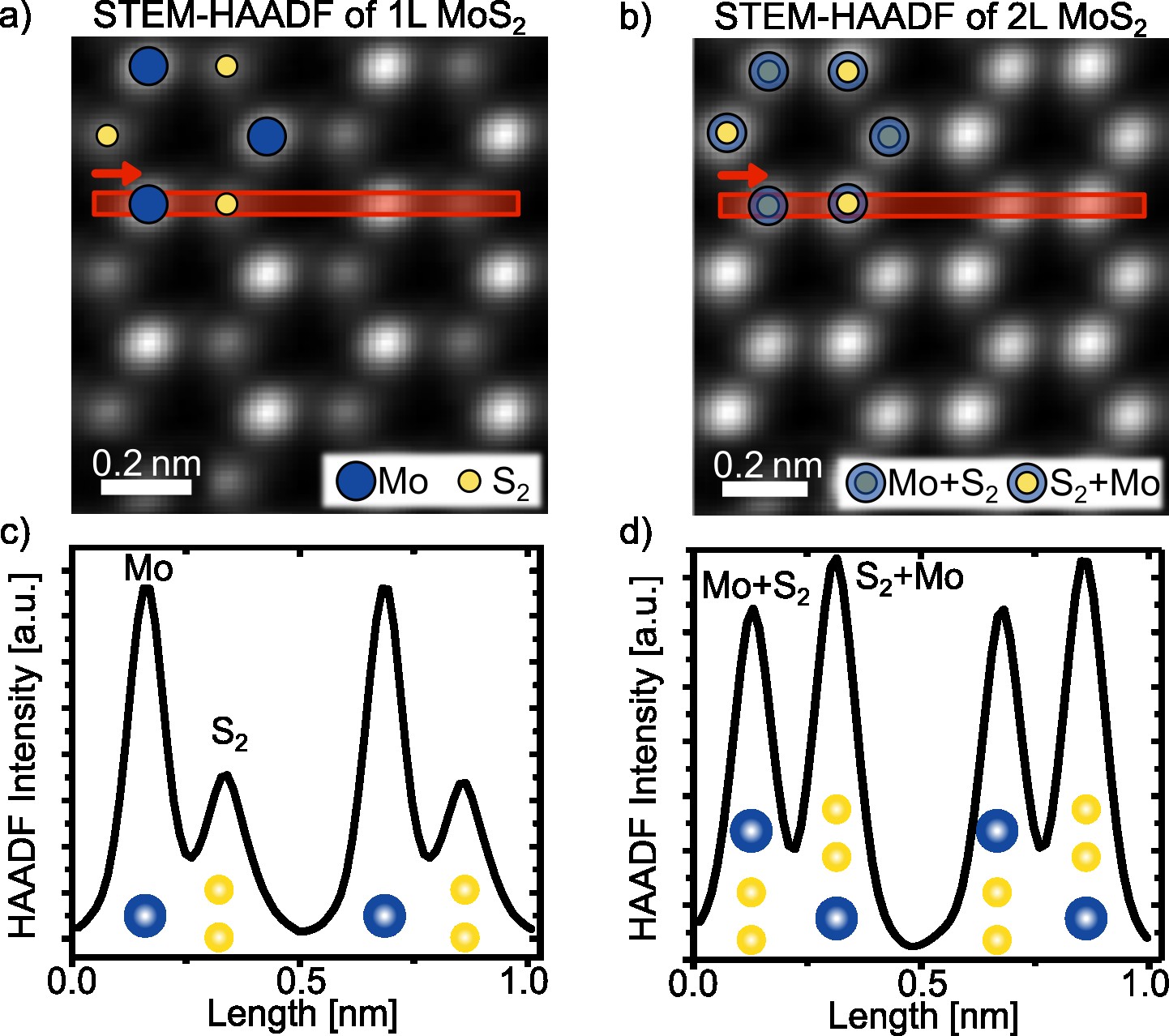}
    \vspace*{-0mm}
    \caption{(a) and (b) STEM-HAADF images obtained in the $[0001]$ zone-axis orientation of 1-ML and 2-ML MoS\textsubscript{2}, respectively. 
    The atomic column positions are marked by the blue and yellow dots for molybdenum and sulfur atomic columns. 
    (c) and (d) Integrated line profiles of the HAADF intensity obtained from the region marked by the red boxes in (a) and (b), respectively.}
    \label{fig:SI-TEM}
\end{figure}

Figure \ref{fig:SI-micrograph} shows an exemplary visible light micrograph of a MoS$_2$ flake on SiO$_2$/Si substrate, where a mono- and a bilayer can be identified by their image contrast.
Using high-resolution STEM-HAADF measurements of free-standing MoS$_2$ mono- and bilayers (see Fig.~\ref{fig:SI-TEM}a and b), the atomic structure as well as the type of atomic columns for the different layers can be characterized by means of the $Z$ contrast in HAADF imaging. 
The image of the monolayer (see Fig.~\ref{fig:SI-TEM}a) reveals the presence of two different atomic columns visible by spots of different brightness.
The differences in intensity are quantified by the integrated line profile shown in Fig.~\ref{fig:SI-TEM}c, taken over two Mo-S$_2$ dumbbells marked by the red box.
Since columns with a higher projected atomic number exhibit higher HAADF image intensity compared to those with a lower projected atomic number, brighter spots corresponds to the Mo ($Z=42$) sublayer, while the darker ones to the two overlapping S ($Z=16$) sublayers. 
The atomic columns with either a Mo atom or two S atoms are indicated in Fig.~\ref{fig:SI-TEM}a by the blue and yellow dots, respectively.
For the bilayer, the HAADF image shows the hexagonal arrangement of atomic columns with comparable HAADF intensities for all atomic columns indicating the presence of the AA’ stacking configuration as this stacking order yields two mixed columns with the same projected atomic number (see Fig.~\ref{fig:SI-TEM}b).

The corresponding integrated line profile, however, reveals the presence of two peaks with slightly different HAADF intensities, see Fig.~\ref{fig:SI-TEM}d.
Thereby, the differences are caused by the stacking sequence of atoms in beam propagation direction and are influenced by the collection angles of the detector. 
Based on literature \cite{Wan2021},\cite{Xia2015}, the atomic columns with lower HAADF intensity in a MoS$_2$ bilayer consist of a Mo atom in the first layer, followed by two S atoms in the second layer. The atomic columns with higher HAADF intensity, on the other hand, consist of two S atoms, followed by one Mo atom.

\begin{figure}[!ttt]
\centering
\includegraphics[width=0.45\textwidth]{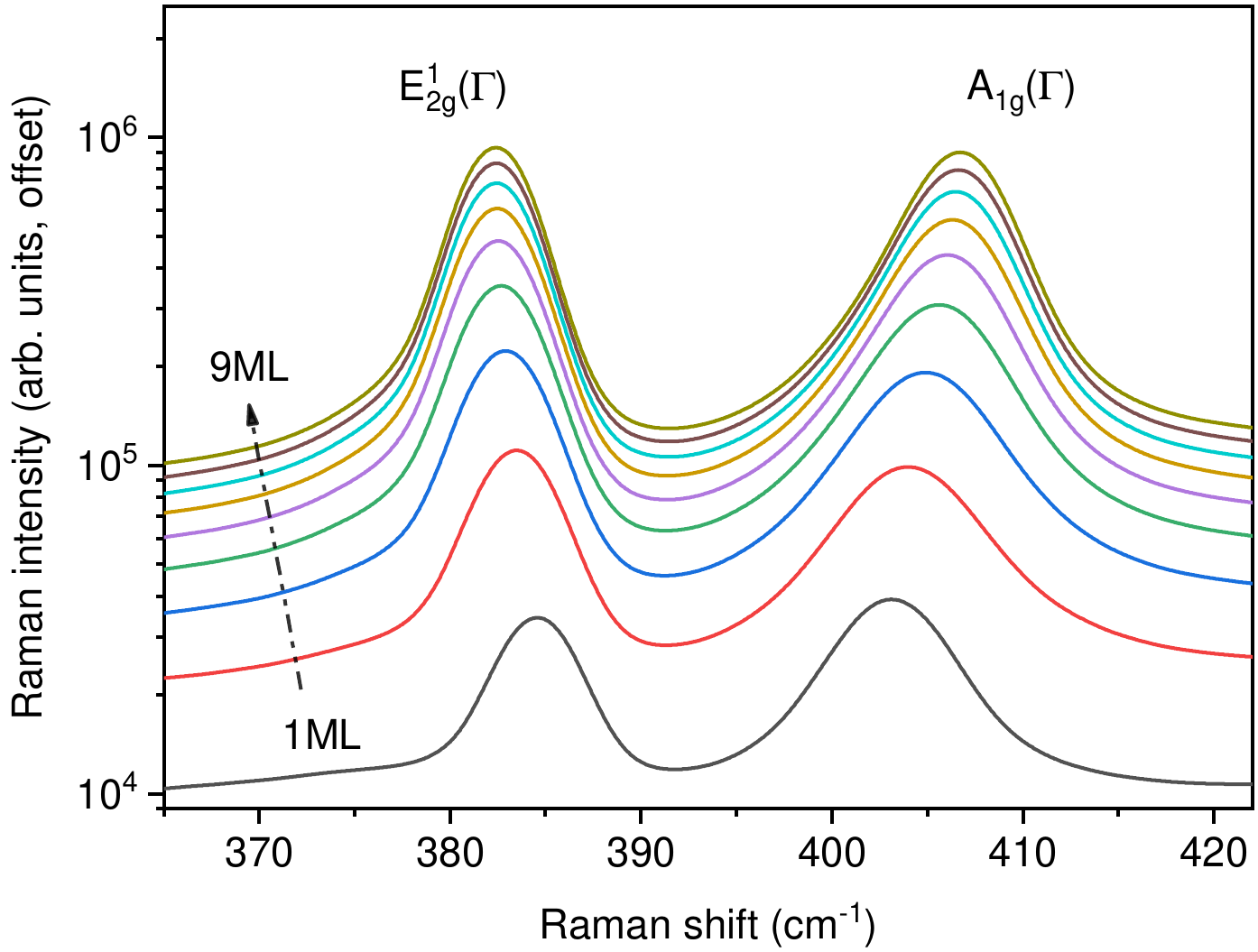}\vfill 
\includegraphics[width=0.45\textwidth]{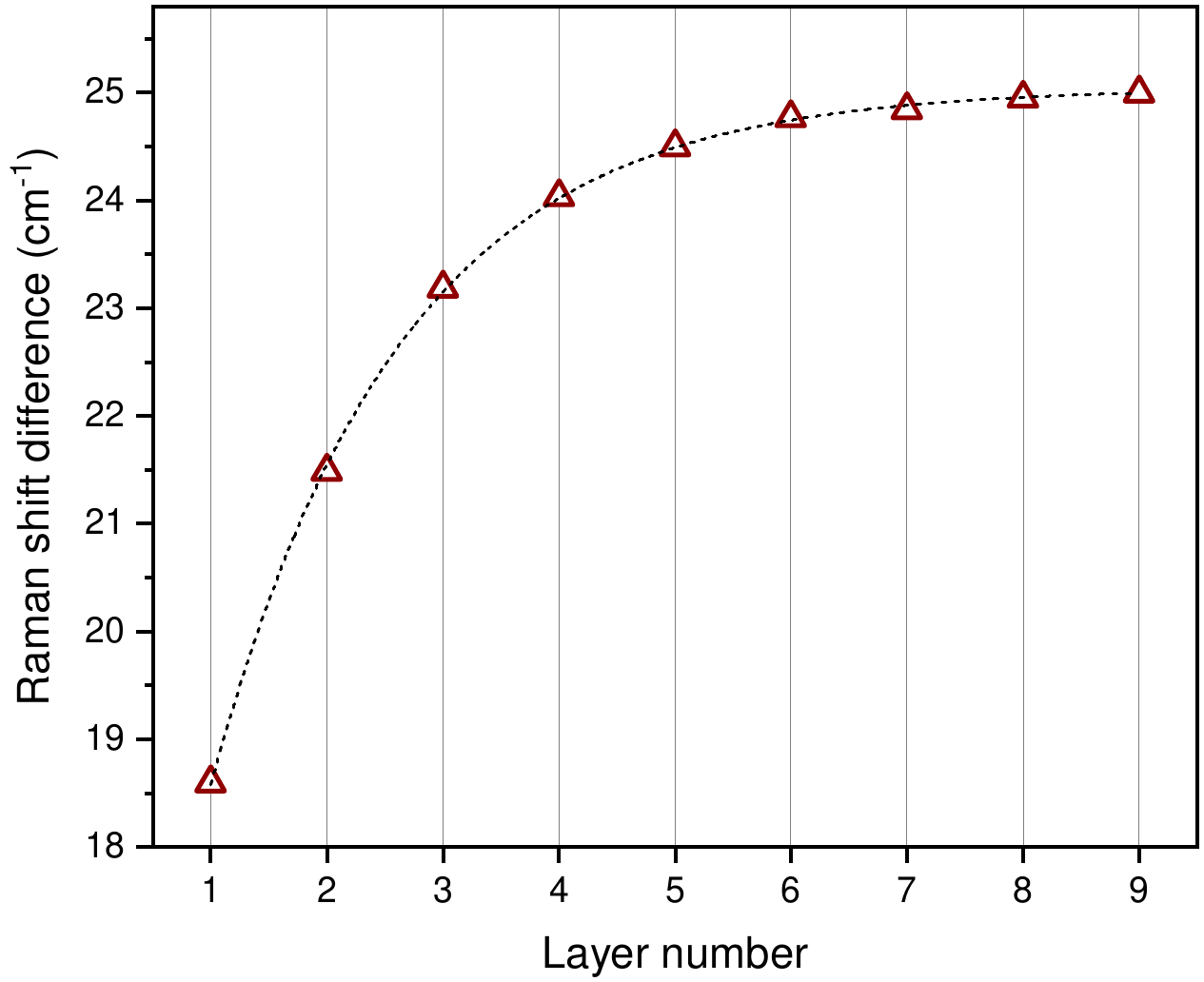}
    \vspace*{-3mm}
    \caption{(Top) Evolution of the Raman spectra of MoS$_{2}$ on SiO$_2$/Si substrate centered around the two main Raman modes E$^{1}_{2g}(\Gamma)$ and $A_{1g}(\Gamma)$, and (bottom) absolute shift difference of the two modes in dependence of the layer number.}
    \label{fig:SI-raman}
\end{figure}

The Raman spectra of MoS$_2$ show two primary phonon modes, \ie\ E$^{1}_{2g}(\Gamma)$ and A$_{1g}(\Gamma)$, which are blue- and red- shifted, respectively, with increasing layer number. \cite{Lee2010AnomalousMoS2}
We find that, compared to the corresponding bulk values, the E$^{1}_{2g}(\Gamma)$ mode is blue-shifted by up to $\approx 2.5$\,cm$^{-1}$, whereas A$_{1g}(\Gamma)$ is red-shifted by up of $\approx 5.0$\,cm$^{-1}$ in the case of the 9-ML system, as evidenced in Fig.~\ref{fig:SI-raman}a.
Although this effect is quite pronounced, to overcome systematical fluctuations, we use the absolute shift difference of the two modes depicted in Fig.~\ref{fig:SI-raman}b rather than their individual values for the analysis. 
Here, we observe that the results of the Raman analysis clearly confirm the estimated layer number.

\begin{figure*}[!ttt]
\centering
\includegraphics[width=0.78\textwidth]{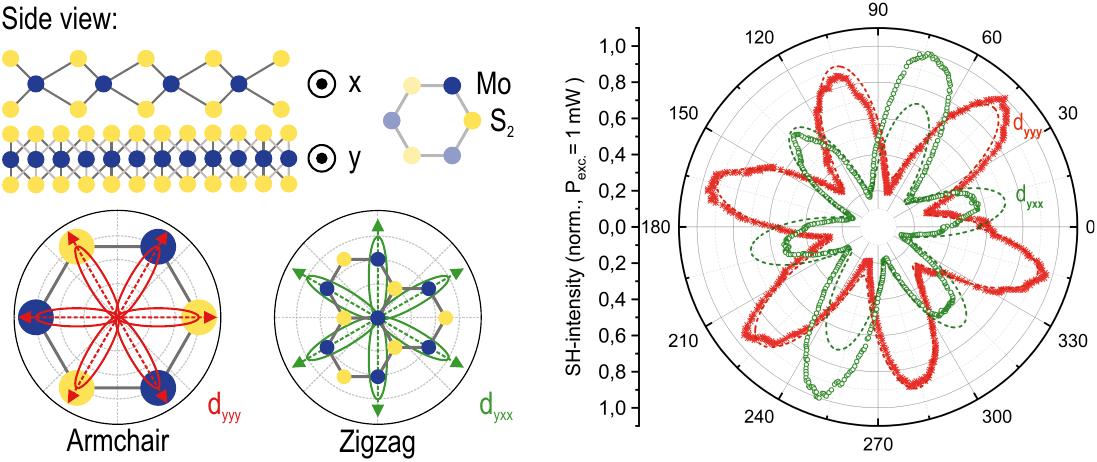}
    \caption{Crystal structure (lhs) and polarization-dependent SH response (rhs) for both d$_{yyy}$ and d$_{yxx}$ tensor elements of the MoS$_2$ monolayer.  
    The red and the green hexagonal patterns are associated with the armchair and zigzag direction, respectively, as depicted in the bottom left corner.
    \label{fig:direction}}
\end{figure*}

MoS$_2$ monolayers belong to the D$_{\mathrm{3h}}$ symmetry group, featuring only one, independent, finite nonlinear tensor element d$_{yyy}=- d_{yxx}=-d_{xxy}=-d_{xyx}$.
The tensor elements d$_{yyy}$ and d$_{yxx}$ are commonly referred to as "armchair" (red) and "zigzag" (green) direction, respectively, see Fig.~\ref{fig:direction}.
Both co-polarized and cross-polarized incident light can induce the SH process, provided that the incident polarization is also aligned with the crystal axes of the material.
Therefore, for free-standing, defect-free MoS$_2$ the two tensor elements (whilst being shifted by 30$^\circ$) are expected to be equivalent in magnitude and to show a perfect, six-fold symmetry, reflecting the crystal structure.
For the polarization-dependent analysis, the excitation and detection polarization are rotated simultaneously, once in parallel orientation (d$_{yyy}$) and once in perpendicular orientation (d$_{yxx}$), measuring the resulting SH intensity for every $0.5^\circ$ rotation.
Although the measurements exhibit the expected six-fold symmetry for both polarizations, see Fig.\ref{fig:direction}, only the d$_{yyy}$ tensor retains the expected uniformity, while the arms of  d$_{yxx}$ tensor appears to be squeezed. 
This occurs most likely due to induced strain during exfoliation along the corresponding crystal axis.
Due to this pronounced asymmetry, we will therefore focus in the following on the (more symmetric) d$_{yyy}$ tensor.

\begin{figure}[!ttt]
\centering
\includegraphics[width=0.48\textwidth]{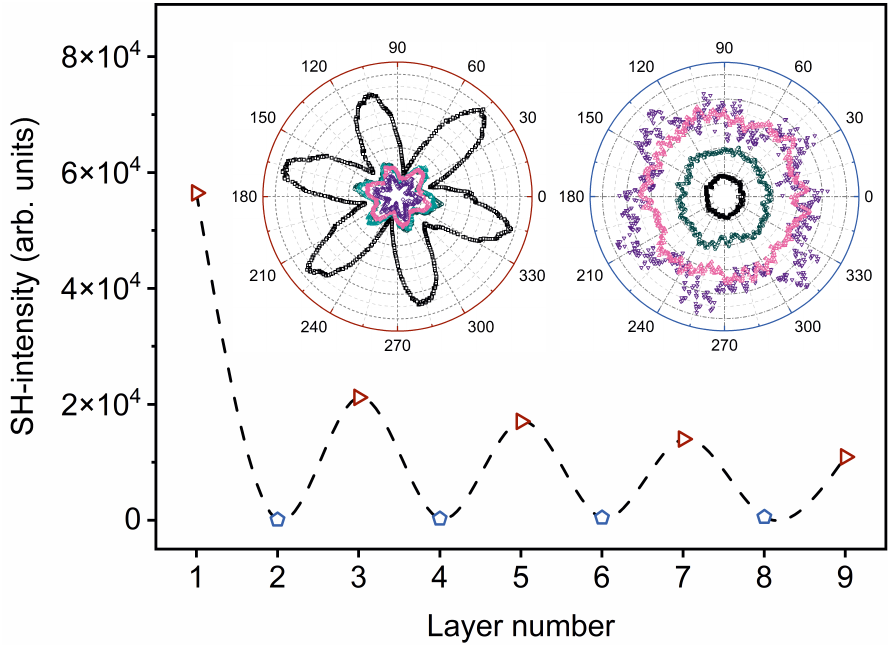} \vfill
\includegraphics[width=0.48\textwidth]{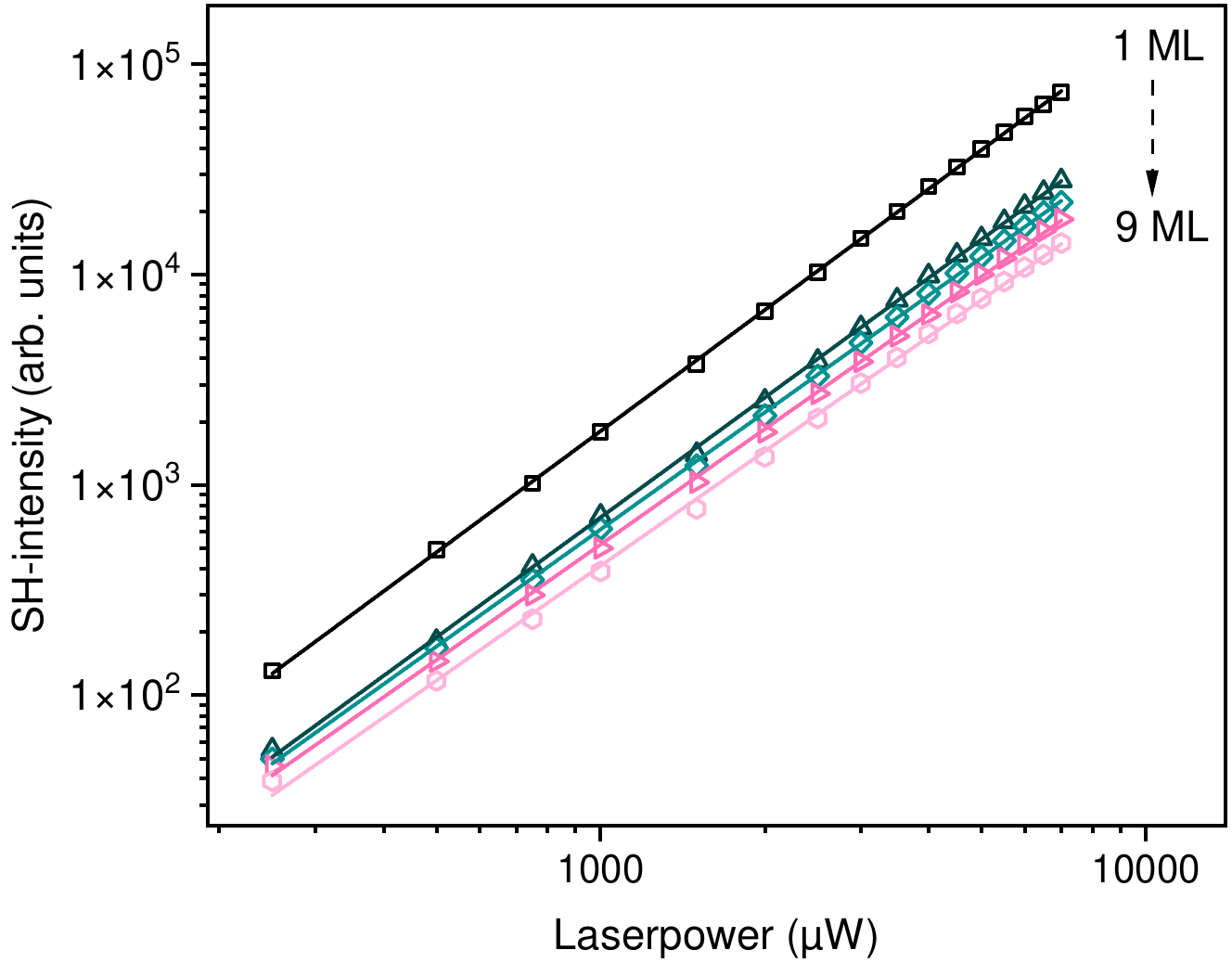}
    \vspace*{-4mm}
    \caption{(Color online) (a) Progression of the SH intensity in dependence of the layer number for a given excitation power and orientation. The measurements conducted on odd-layer MoS$_2$ reveal that (upon calibration) SH microscopy offers a non-invasive method to determine the layer thickness of few-layer MoS$_2$ films.
    The insets show polarization scans for odd- (left) and even-layer (right) MoS$_2$;
    (b)  Laser-power-dependent measurements conducted on odd-layer MoS$_2$ in a double logarithmic display. The slopes of the linear fits range from $1.914$ (1 ML) to $1.811$ (9 ML) in a consecutive decrease, verifying the expected parabolic behavior.
    }   
    \label{fig:layer_thickness_trend}
\end{figure}

Beyond the monolayer, few-layer MoS$_2$ can be stacked in various ways. Here, the MoS$_2$ samples are obtained from a 2H-stacked bulk crystal via mechanical exfoliation.
Therefore, the few-layer configurations retain the same stacking as in the bulk.
Consequently, inversion symmetry is broken and preserved for odd-layer and even-layer systems, respectively.
For this, only the former are expected to yield a significant SH signal.
Nevertheless, as shown in the right inset in Fig.~\ref{fig:layer_thickness_trend}a, a modest signal is resolved also for even-layer samples.
We attribute these effects to structural disturbances like surface defects, interface effects with the substrate \cite{trolle2014} or to strain induced during the fabrication process or by the underlying substrate, since the effect is most pronounced for lower layer thicknesses.

The polarization-dependent measurements of odd-layer MoS$_2$, on the other hand, reflect the six-fold symmetry of the crystal structure as the monolayer, see right inset in Fig.~\ref{fig:layer_thickness_trend}a. 
In addition, power-dependent measurements (see Fig.~\ref{fig:layer_thickness_trend}b) reveal that the by far highest signal is obtained for the monolayer, while for thicker samples the SH intensity is reduced steadily.
However, the slope of the fitting curves deviates slightly from the expected value of two.
This discrepancy could be caused by the presence of the substrate, or (considering that the deviation increases with the layer number) to in-sample absorptions of the SH radiation.
Due to this pronounced effect, upon calibration SH microscopy represents a powerful and efficient tool to determine the thickness of few-layer MoS$_2$ \cite{kumar2013}.

\begin{figure*}[!ttt]
\centering
\includegraphics[width=0.95\textwidth]{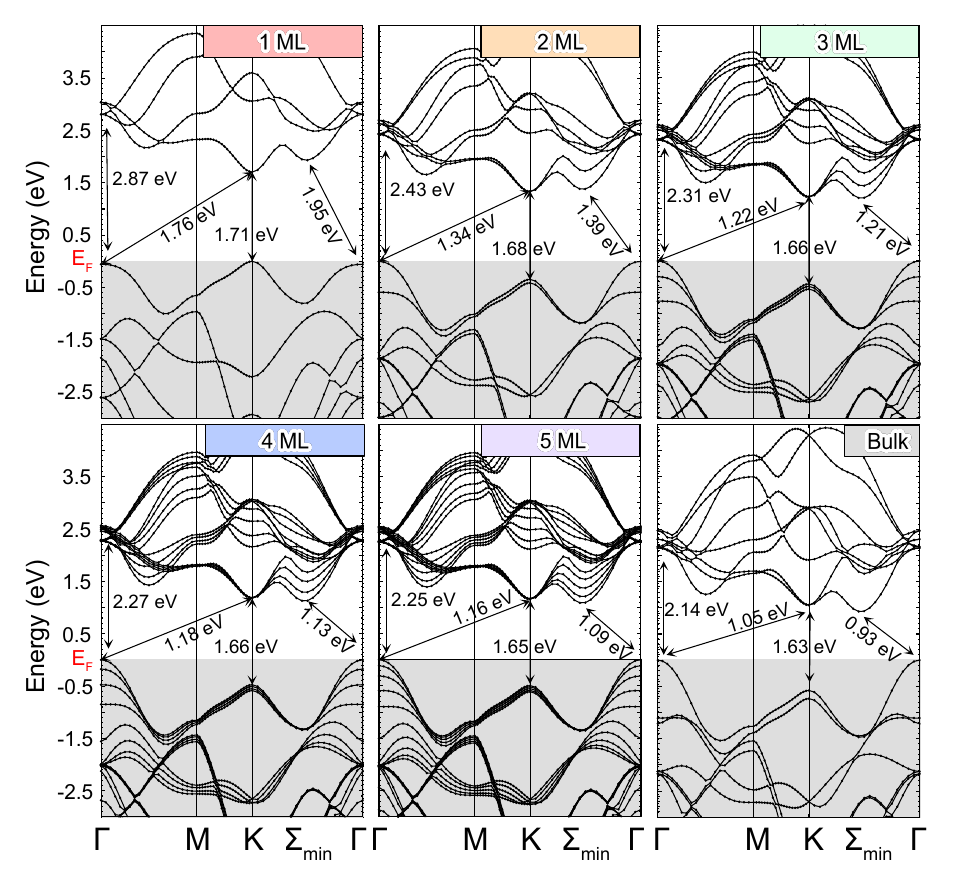}
    \vspace*{-5mm}
    \caption{(Color online) Calculated PBE-DFT band structure of few-layer and bulk MoS$_2$ along the path $\Gamma \to M \to K \to \Gamma$.
    The zero line is set to the Fermi-level position ($E_F$).}
    \label{fig:SI-geometry}
\end{figure*}

SH intensities of layer configurations with odd number parity (for the given excitation energy of $780$\,nm and an excitation power of $6$\,mW) are found to scale approximately with $\nicefrac{1}{n}$ ($n$ indicating the layer number), see Fig.~\ref{fig:layer_thickness_trend}a.
Similarly, also Malard \etal\ \cite{malard2013} reported a strong reduction of the SH signal upon the transition from the 1-ML to the 3-ML system. They attributed the differences in the two functions to changes in the band structure.
However, while the strong SH-intensity loss from the mono- to the trilayer could be related to the transition from a direct to indirect gap, the reason for the decreasing SH intensity of higher-layer samples cannot be purely explained with changes of the band structure:
The DFT band-structure calculations reported in Fig.~\ref{fig:SI-geometry}, in fact, reveal that the differences between the one of the trilayer and the 5-ML system are only minor so that they would only explain slight deviations in the SH intensity.
More precisely, the direct gap at $\Gamma$ steadily decreases with an increasing number of layers starting from the value of $2.87$\,eV in the case of the monolayer and reaching the value of $2.25$\,eV in the case of the five-layer system.
The differences for the direct gap at the $K$ point, on the other hand, are even less pronounced in all the investigated cases.
This was attributed to a strong perpendicular quantum-confinement, which only affects the size of the indirect gap \cite{mak2010}.

\begin{figure*}[!ttt]
\centering
\includegraphics[width=0.98\textwidth]{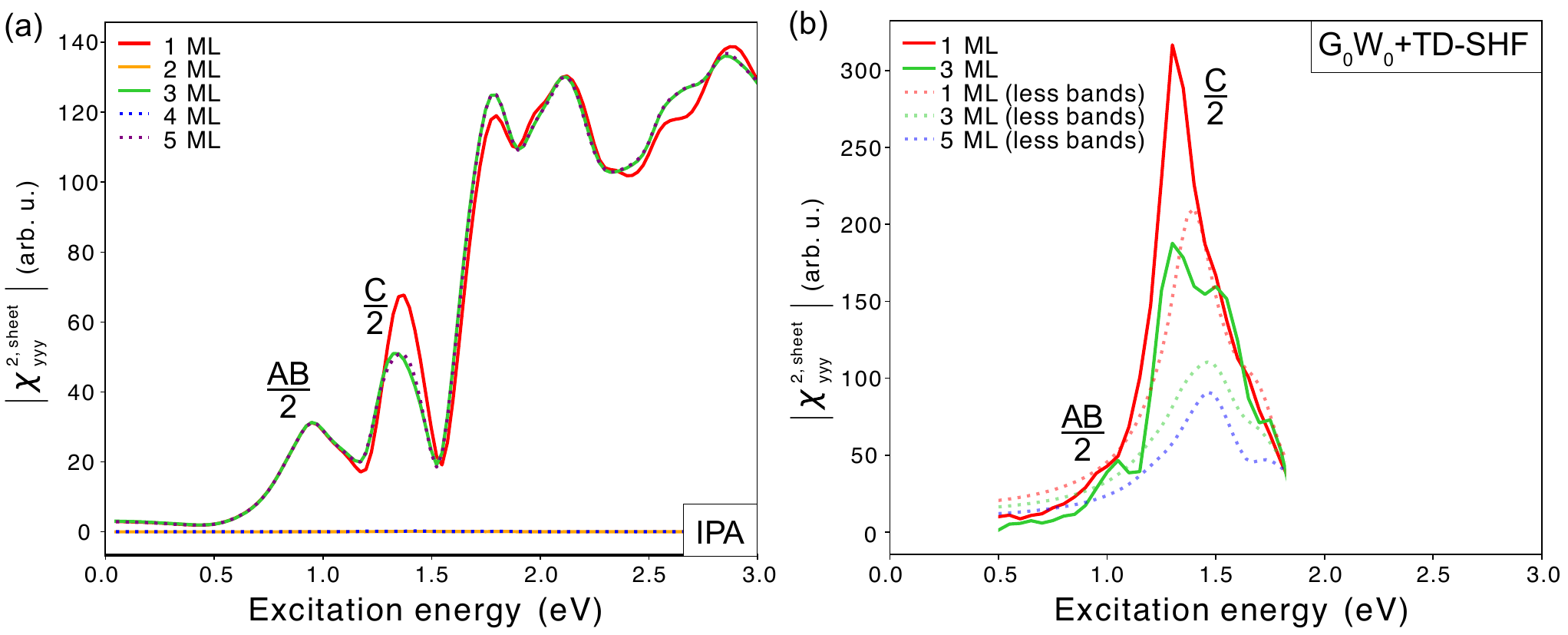}
    \vspace*{-3mm}
    \caption{(a) IPA and (b) \textit{G$_0$W$_0$}+TD-SHF second-order 2D susceptibility tensor (\ie\ $\left|\chi^{2, \mathrm{sheet}}_{yyy}\right|$) of few-layer MoS$_2$ in dependence of the excitation energy.}
    \label{fig:optics}
\end{figure*}

This statement is reinforced by our simulated IPA susceptibility depicted in Fig.~\ref{fig:optics}a in dependence of the excitation energy.
Note that in order to compare the intensities of spectra of supercells of different size, we determine the sheet susceptibility $\left|\chi^{2, \mathrm{sheet}}_{yyy}\right|$ by multiplying the intensity of the computed spectra with the height of the corresponding supercell \cite{xiao2022}.
As expected, due to inversion symmetry, the $\left|\chi^{2, \mathrm{sheet}}_{yyy}\right|$ tensor of even-layer MoS$_2$ films is zero, irrespective of the excitation energy.
For odd-layer systems, on the other hand, we obtain a good description of the $\nicefrac{C}{2}$ peak, see Fig. \ref{fig:optics}, (centered at an excitation energy of about $1.4$\,eV \cite{malard2013}) due to the good approximation of the band gap achieved already for the PBE functional.
Interestingly, however, the exact same function for the SH response of the 3-ML and the 5-ML system is obtained, which is furthermore very similar to the one of the monolayer.
More precisely, the $\nicefrac{AB}{2}$ peaks of all three systems coincide, while the $\nicefrac{C}{2}$ peak of the monolayer is slightly higher in intensity compared to the one of the 3-/5-ML system.
This behavior can be easily explained by referring to the band structures of the systems:
The former is caused by direct transitions at the $K$ point, where the gap is almost constant irrespective of the layer number.
The latter, on the other hand, by transitions in the layer-number-dependent region between the $\Gamma$ and the $K$ point, also see Fig.~\ref{fig:SI-geometry}.
This also explains the slight shift in the position of the $\nicefrac{C}{2}$ compared to the $\nicefrac{AB}{2}$ peak.
So, our IPA spectra do not reflect the trends resolved in the experiment, and the layer-dependent intensity loss can
only to a small extent be attributed to modifications in the band structure.

Excitonic effects, however, have been known already to yield to a strong redistribution of the intensities \cite{trolle2014, lumenMoS2} for the monolayer. Therefore, they can be expected to also have a strong impact on few-layer systems
So, we recalculate the $\left|\chi^{2, \mathrm{sheet}}_{yyy}\right|$ susceptibility upon the explicit inclusion of quasiparticle and excitonic effects, \ie\ via the $\textit{G$_0$W$_0$}+$TD-SHF formalism.
We focus our investigation on odd layered system, since even layered structures are expected to exhibit no reasonable nonlinear behavior and computational costs can be reduced this way.
The so-obtained spectra are depicted in Fig.~\ref{fig:optics}b:
Compared to the IPA, the $\textit{G$_0$W$_0$}+$TD-SHF $\left|\chi^{2, \mathrm{sheet}}_{yyy}\right|$ susceptibility of the monolayer shows a strong redistribution of the oscillator strengths upon the inclusion of excitonic effects, yielding a very strong increase of the SHG as already reported in Refs.~\citenum{lumenMoS2, trolle2014}.
Similarly, also the features of the 3-ML system are risen in intensity and now show clear differences in line shape compared to the monolayer:
Not only is the main peak intensity notably lowered compared to the one of the monolayer, but a shoulder on the right side of the peak appears, which is in very good agreement with the experiment \cite{malard2013}. 
We expect the $\left|\chi^{2, \mathrm{sheet}}_{yyy}\right|$ susceptibility of the 5-ML system to be even more reduced in intensity compared to the 3-ML one, as shown by the dotted lines in Fig.~\ref{fig:optics}b.
These dotted lines refer to calculations with fewer bands (i.e., only three valence and two conduction bands per monolayer), in order to make larger systems computationally accessible.
The intensity decrease with increasing film thickness, which is exclusively observed in calculations including excitonic effects, points to exciton delocalization as mechanism for the SHG intensity lowering. 
Indeed, while for the monolayer and bilayer intra-excitons are confined to the surfaces, for thicker samples, quasi-inter-layer excitons distribute throughout the layers \cite{das2019}, suggesting a surface-exciton enhancement of the SH response of the material.
Furthermore, it has been already discussed for linear optical properties that the excitonic binding energy for 2D materials \cite{olsen2016} depends on the polarizability, diminishing for higher layer numbers, which is also expected to influence the nonlinear properties.

In addition, we suggest that losses caused by in-sample absorption of the generated SH radiation might even further impact the SH intensity resolved in the experiment:
It has been shown \cite{MoS2BSE} that MoS$_2$ single and double ML show different linear absorption signatures in the region around 3\,eV due to strong bound excitons.
These could therefore also impact the absorption properties of films in dependence on the number of layers.

\section{Conclusions}

In this study we have thoroughly investigated the nonlinear optical properties of pristine MoS$_2$ few-layer systems.
For this, we fabricated and characterized structures of up to nine layer thickness, which were observed to be distinguishable via their nonlinear response:
While a spurious SH signal resolved for even-layer systems can be attributed to a locally broken inversion, for odd-layer samples a clear trend of reduction in the SH intensity for increasing sample thicknesses is observed. 
More precisely, for the excitation energy of $780$\,nm, the SH intensity scales roughly with $\nicefrac{1}{n}$ with $n$ indicating the layer number.
This effect cannot be purely attributed to band-structure effects.
In fact, only upon the explicit inclusion of excitonic effects, the experimentally resolved reduction in SH intensity is reproduced by our simulations, while their neglection yields  closely related if not even identical
$\left|\chi^{2, \mathrm{sheet}}_{yyy}\right|$ in all investigated cases.
Thus, our data support the scenario in which the reduction of SH intensity is related to the spatial localization of excitons and in-sample absorption of the generated SH radiation, which is important to consider when exploiting effects that include excitonic features.
We further suggest that the effects here discussed for MoS$_2$ can likely be applied to similar materials within the TMDC group.

\section*{Acknowledgements}
This work is supported by the Deutsche Forschungsgemeinschaft (German Research Foundation, DPG) through the transregional collaborative research center TRR 142/3-2025 (Project No.~231447078) and by the European Research Council (ERC) under the European Union’s Horizon 2020 research and innovation program (LiNQs, 101042672). K.D.J. acknowledges funding from the Ministry of Culture and Science of North Rhine-Westphalia for the Institute for Photonic Quantum Systems (PhoQS).
We also thank the Paderborn Center for Parallel Computing (PC$^2$) for grants of high-performance computational time.

\bibliography{paper.bib}

\end{document}